\documentclass[aps,prl,twocolumn,groupedaddress]{revtex4}

\begin{document}

\date{\today }

\textbf{Comment on "Interaction-driven Relaxation of Two-level Systems in
Glasses"} In a series of our previous publications a scaling theory of the
interaction stimulated relaxation (ISR) in ensemble of two-level-systems
(TLS) coupled as $U_{0}/R^{3}$ has been proposed \cite{seeEX1998} to
describe the linear temperature relaxation rate $\tau _{1}^{-1}=c\cdot
\left( P_{0}U_{0}\right) ^{3}T$ observed in certain ultra-low temperature
experiments \cite{G-Bayer1988, P-Esq1992, S Rogge -1996}. Here $P_{0}$
characterizes the density of TLS\ per unit energy interval and the $c$ is a
numerical pre-factor which cannot be found within our scaling approach. To
obtain a quantitative agreement between the theory and the experiments, the
pre-factor should be of the order of $10^{3}.$

In a recent Letter \cite{wuerger} D. Bodea et al. have attempted to attain
the quantitative agreement with the experiment suggesting an alternative
scenario for the ISR of TLS. Their approach is based on the resonant triples
of TLS. First the authors estimated a coupling amplitude for a triple $J$ as%
\begin{equation}
J^{2}=\overline{u}_{i}^{2}J_{ik}^{2}+\overline{u}_{j}^{2}J_{jk}^{2}.
\label{1}
\end{equation}%
Here $\bar{u}=\Delta /E,$ with $\Delta $ being the asymmetry energy, $E$ is
the energy splitting and $J_{ik}=U_{0}/R_{ik}^{3}.$This estimate has been
served as a basis for establishing the energy delocalization mechanism and
for calculating the TLS\ relaxation rate induced by this mechanism. We
cannot agree with the results of Ref. \cite{wuerger}, since, in our opinion,
the triple transition amplitude (\ref{1}) is qualitatively overestimated. It
seems evident that any many-TLS transition amplitude, in particular the
triple amplitude Eq.(\ref{1}), should decrease when one of the TLS moves off
from the others. Yet, it follows from Eq. (\ref{1}) that, if $R_{ij}\approx
R_{jk}\gg R_{ik},$ the triple transition amplitude becomes independent of
the $R_{jk}$ instead of vanishing.

Using second order perturbation theory in TLS\ interaction one can derive
the correct triple TLS coupling amplitude $J$ in the form \cite{condmat}%
\begin{equation}
\begin{array}{c}
J_{ijk}\approx -\frac{\Delta _{0i}\Delta _{0j}\Delta _{0k}}{8E_{i}E_{j}E_{k}}%
\times  \\
\times \left[ -\frac{U_{ij}U_{ik}\Delta _{i}}{E_{j}E_{k}}-\frac{%
U_{ij}U_{jk}\Delta _{j}}{E_{i}E_{k}}+\frac{U_{ik}U_{jk}\Delta _{k}}{%
E_{i}E_{j}}\right]
\end{array}
\label{2}
\end{equation}%
significantly different from Eq. (\ref{1}). Using this expression one can
readily show that the probability to form a resonant triple is much smaller
than unity so the triples, like pairs, can not result in any relaxation. To
our mind, these reasonings disprove the conclusions of Ref. \cite{wuerger}.

We understand that the lack of the pre-factor of the order of $10^{3}$ in
our theory can alert the reader. However, obtaining the very underestimated
relaxation rate $\tau _{1}^{-1}$ is a feature of the scaling approach, where
the delocalization criterion is about one order stronger than that obtained
within the consistent microscopical approach cf. Ref. \cite{pwanderson}.
Therewith, in our scaling approach \cite{seeEX1998} we properly consider
only resonant interaction when energy level spacing is smaller than the
transition amplitude coupling these levels. The contribution of nonresonant
interaction cannot qualitatively change the scenario of the many- body ISR\
relaxation \cite{seeEX1998}, but might introduce an extra large numerical
factor into our result. A similar effect seems to be observed in the system
of interacting electrons in disordered wires \cite{mirlin} where the
renormalization of the dimensionless disorder parameter $\alpha $ takes
place ($\alpha \rightarrow \alpha \left\vert \ln \alpha \right\vert $) when
estimating the delocalization criterion. In our case, this could result in
the substitution $P_{0}U_{0}\rightarrow P_{0}U_{0}\left\vert \ln \left(
P_{0}U_{0}\right) \right\vert \approx 10P_{0}U_{0}$ and reconcile the theory
and the experiment.

A large pre-factor is known to appear when integrate over the phase volumes
of many dimensions simultaneously. In particular, the accurate estimate of
the swing of spectral diffusion in glasses found in Ref. \cite%
{black-halperin} is about $10^{3}$ larger than that obtained within the
scaling approach \cite{seeEX1998}. Also, by the same reason, the quantum
hopping diffusion coefficient of light impurities in crystals acquire a huge
pre-factor $10^{7}$ \cite{Yu.Kagan}.

Thus, the scenario of ISR induced by a long range $R^{-3}$-interaction
proposed in Ref. \cite{seeEX1998} results in the most fast relaxation
channel. Within the range of applicability of the scaling approach, the
theory correctly predicts the linear temperature behavior for the TLS\
relaxation rate. One method of experimental verification of our approach is
measuring the relaxation rate as a function of the density of TLS $%
P_{0}U_{0}.$

A. L. Burin, I. Ya. Polishchuk,

Department of Chemistry, Tulane University, New Orleans, LA 70118, USA

Max-Planck-Institut f\"{u}r Physik komplexer Systeme, D-01187 Dresden,
Germany

\end{document}